\documentclass[11pt]{article}

\usepackage{amsmath,amssymb} 

\newtheorem{remark}{Remark}[section]

\newtheorem{theorem}{Theorem}[section]

\newtheorem{proposition}{Proposition}[section]

\newtheorem{corollary}{Corollary}[section]

\newcommand{\proof}{ $\triangleright$\quad }
\newcommand{\qed}{\hfill $\triangleleft$}

\begin{document}

\title{Functional derivatives, Schr\"{o}dinger equations, and  Feynman integration}

\author{Alexander  Dynin\\
\textit {\small Department of Mathematics, Ohio State University}\\
\textit {\small Columbus, OH 43210, USA}, \texttt{\small dynin@math.ohio-state.edu}}

\maketitle

\begin{abstract}
  Schr\"{o}dinger equations  in  \emph{functional derivatives} are solved via  quantized Galerkin  limit of antinormal functional Feynman  integrals for Schr\"{o}dinger equations in \emph{partial derivatives}.
\end{abstract}

\medskip



\bigskip
\begin{quotation}
\noindent \textsl{Thus there arise an interpretation of the second quantization problems  
as quantum mechanics problems with infinitely many degrees of freedom and a natural desire 
to approximate  these problems via problems with finite, but large, number of degrees of freedom.}

F.A. Berezin (\cite{Berezin}, Introduction).
\end{quotation}

\normalsize
\section*{Introduction}

The very  first  stationary  functional derivatives  Schr\"{o}dinger
equation  was introduced in 1928 by  P. Jordan and W. Pauli 
( Zur Quantumelectrodynamik ladungsfreier Felder,  
Zeitung f\"{u}r Physik, Vol. 47): 

For  wave functionals $F(\phi(x))$
of massless scalar fields $\phi(x), x\in\mathbb{R}$,
\begin{equation}
-\left(\frac{\hbar}{4\pi}\right)^2\int\!dx\, \left[
\frac{\delta^2}{\delta \phi(x)^2}
+ c^2\left(\frac{d\phi(x)}{d x}\right)^2\right]F(\phi(x))=\lambda F(\phi(x)).
\end{equation}
There was a vivid discussion of "Volterra mathematics"  between W. Heisenberg, P. Jordan, and W. Pauli. However until now there has been no  sound mathematical progress in solution of such equations. Pertrurbation and  lattice approximations do not converge in  meaningful examples.

Moreover, according to  P. Dirac ( ``Lectures on quantum field theory" Yeshiva University, N.Y. 1966, 
Section  ``Relationship of the Heinseberg and  Schr\"{o}dinger Pictures"),
\begin{quotation}
\textsl{
The interactions that are physically important in quantum field theory are so violent 
that they will knock  any Schr\"{o}dinger state vector out of Hilbert space  
in the shortest possible time interval.}

\textsl {[...] It is better to  abandon all attempts at using the Schr\"{o}dinger picture 
with these Hamiltonians.}

\textsl {[...] I don't want to assert that the Schr\"{o}dinger picture will not come back. 
In fact, there are so many beautiful things about it that I have the feeling in 
the back of my mind that it ought to come back. I am really loath to have to give it up.}
\end{quotation}

Heisenberg partial derivatives equations for  interacting  quantized fields  are non-linear. In contrast, Schr\"{o}dinger  equations  for states  are linear. Presumably they may be solved by well developed   Hilbert space methods.

Unfortunately, in the second quantization formalism,  a  ``violent"   Schr\"{o}dinger  operator is not densely defined in the
   Fock space (see \cite{RS}, vol. II, Chapter X). For the sake of operator methods one needs  to apply cutoffs.  

This article proposes a rigorous mathematical theory of  Schr\"{o}dinger functional differential operators with combined ultraviolet and infrared cutoffs: 
\begin{itemize}
\item Section 1 is a  convenient review of infinite dimensional distributions. 
\item   Section  2 begins with  a rigorous  treatment of functional derivative operators.
Theorem 2.2 asserts a lower  bound for  cutoff   Hamiltonian functional derivative operators defined by classical Hamiltonians bounded from below. 
The  coherent states matrix elements of  the corresponding evolution operators are    given the form of  antinormal functional Feynman  integral  (Theorem 2.3).

\item  Section 3   introduces a quantized infinite dimensional Galerkin approximation of cutoff   functional derivative equations  by  partial derivative equations. shows  that  
this Feynman integral is a  double limit of  finite dimensional Gaussian integrals  (Theorem 3.1). Thus we have a convergent computational scheme in pseudo-Euclidean space,
a viable alternative to lattice approximations. 
\end{itemize}

The original antinormal   Feynman  integral, based on Chernoff's product formula,   was introduced  by
J. Klauder and B. Scagerstam (see  \cite{KS}, page 69). Another, based on an infinite-dimensional symbolic calculus, has been used in \cite{D} to solve non-cutoff functional Schr\"{o}dinger equations  with integrable infinite-dimensional Hamiltonians. 

Here the functional Feyman integral   is rigorously defined  as a  limit of   Klauder-Scagerstam integrals associated  with  approximating finite-dimensional Hamiltonians.

\medskip
\emph{In the text the triangles  $\triangleright$ and $\triangleleft$ mark  the beginning and the end of a proof.}

\section{Review of infinite-dimensional distributions}

\subsection{Bosonic Fock representations}
Let   $\mathcal{H}$  be a complex (separable) Hilbert  $*$-space  with a given complex conjugate isometric involution 
$\phi\rightarrow \phi ^{*}$. 
 The  $*$-subspaces of $\mathcal{H} $ are invariant under the conjugation.
 
  The Hermitian inner product of $\alpha$ and $\beta$ is denoted
  $\alpha^{*}\beta$.  It is complex conjugate linear in   $\alpha^{*}$ and  linear in $\beta$.

\smallskip 
The \emph{real part }  of $\mathcal{H}$ is the real Hilbert subspace
$\Re\mathcal{H}=\{\rho\in\mathcal{H}: \rho^{*}=\rho\}$, and the \emph{imaginary part}  
of $\mathcal{H}$ is the real Hilbert subspace $\Im\mathcal{H}=\{i\pi\in\mathcal{H}: 
\quad \pi\in\Re\mathcal{H}\}$ .

Since any $\phi=\rho+i\pi$ with $\rho=(\phi+\phi^{*})/2,\  \pi=(\phi-\phi^{*})/2i$, the $*$-space
is the direct orthogonal sum of the real part $\Re\mathcal{H}$ and the imaginary part $\Im\mathcal{H}$.  
Along with $\phi^{*}=\rho-i\pi$ this implies that  a choice of the involution $*$ is uniquely 
defined by  the choice of the  real part  $\Re\mathcal{H}$.

\smallskip
An operator $o$ in $\mathcal{H}$ is \emph{real} if it commutes with
the involution $*$.

\smallskip
Let  $\mathcal{H}^{*}$ denote the antidual Hilbert space  of $\mathcal{H}$ with 
respect to the Hermitian form  $\phi^{*}\psi$.  

\smallskip
The Hilbert space $\mathcal{H}\times\mathcal{H}^{*}$  carries
the  conjugation $(\alpha,\beta^{*})=(\beta,\alpha^{*})$. The
corresponding  real part $\mathcal{R}$
is the \emph{antidiagonal} $\{(\phi,\phi^{*}):\phi\in\mathcal{H}\}$.
The isometry $\phi\mapsto (1/\sqrt{2})(\phi,\phi^{*})$ is
a representation of $\mathcal{H}$ as a real Hilbert space.

\medskip
A  \emph{Fock representation of bosonic canonical commutation relations}   
over $\mathcal{H}$ is  described by 
\begin{enumerate}
 \item
 A Fock Hilbert $*$-space $\mathcal{F}=\mathcal{F}(\mathcal{H})$;
\item 
Two families of  \emph{creators} $\mathcal{F}^{+}(\phi)$ and 
 \emph{annihilators} $\mathcal{F}^{-}(\phi)$ which are linear unbounded operators
 in $\mathcal{F}$ with a common invariant dense $*$-domain
 $\mathcal{P}$ in $\mathcal{F}$  such that  $\mathcal{F}^{+}(\phi)$
 and  $\mathcal{F}^{-}(\phi)$  are complex  linear with  respect to $\phi\in\mathcal{H}$ 
and the  Hermitian adjoint 
$ [\mathcal{F}^{+}(\phi)]^{\dagger}=\mathcal{F}^{-}(\phi^{*})$
 \item  
 The unit \emph{vacuum}  vector  $F_{0}$  in $\Re\mathcal{P}$.
 \item
$\mathcal{F}^{-}(\phi)F_{0}=0$ for any $\phi$; and $\mathcal{P}$
 the linear span of   the \emph{power} Fock vectors
 \begin{equation} 
 \mathcal{F}^{+}(\phi^{*})^{n}F_{0},\  \phi\in \mathcal{H},\  n=0,1,2,... .
 \end{equation}
   \item
 The commutators of creators and annihilators  and satisfy the 
 \emph{ canonical Fock commutation relations}  on  $\mathcal{P}$:
\begin{equation}
[\mathcal{F}^{-}(\alpha^{*}), \mathcal{F}^{+}(\beta)]=\alpha^{*}\beta,\quad
[\mathcal{F}^{+}(\alpha), \mathcal{F}^{+}(\beta)]= 0 
=[\mathcal{F}^{-}(\alpha), \mathcal{F}^{-}(\beta)].
\end{equation}
\end{enumerate}
The polarization formula
\begin{equation}
\mathcal{F}^{+}(\phi_{1})\ldots \mathcal{F}^{+}(\phi_{n})F_{0}=\frac{1}{2^{n}n!}\sum_{\sigma_{j}} \sigma_{1}\ldots\sigma_{n}
\mathcal{F}^{+}(\sigma_{1}\phi_{1}+\ldots+
\sigma_{n}\phi_{n})^{n}F_{0},
\end{equation}
where  $2^{n}$ coefficients $\sigma_{j}\in\{1,-1\},\ j=1,\ldots,n$,
shows that  $\mathcal{P}$ is the compex span  of the product Fock vectors
 $\mathcal{F}^{+}(\phi_{1})\ldots \mathcal{F}^{+}(\phi_{n})F_{0}$.

\begin{remark}
For  a given $\mathcal{H}$ all \emph{Fock representations}
$(\mathcal{F}, \ F_{0}, \ \mathcal{F}^{+}, \mathcal{F}^{-})$ are  unitary  equivalent.
\end{remark} 
The \emph{Segal functor}  $\Gamma$ (see \cite{BSZ}, Chapter I) assigns to an operator 
$o$,  with a  dense domain $\mathcal{H}'$ in $\mathcal{H}$,
an operator $\mathcal{F}(o)$ in $\mathcal{F}$, with the  dense domain $\mathcal{P}'$,  
spanned by $\mathcal{F}^{+}(\phi')^{n},\ \phi'\in\mathcal{H}',  n=1,2...$ such that 
\begin{equation}
\mathcal{F}(o)F_{0}=F_{0},\ 
\mathcal{F}(o)[\mathcal{F}^{+}(\phi')^{n}F_{0}]=\mathcal{F}^{+}(o\phi')^{n}F_{0}.
\end{equation}
Then
\begin{itemize}
\item
If  $o_{2}o_{1}$ exists on a dense domain in $\mathcal{H}$,  then
$\mathcal{F}(o_{2}o_{1})=\mathcal{F}(o_{1})\mathcal{F}(o_{2})$.
\item
$\mathcal{F}(1)=1,\ \mathcal{F}(o^{-1})=\mathcal{F}(o)^{-1},\
\mathcal{F}(o^{\dag})=\mathcal{F}(o)^{\dag}$. 
\item
If $o$ is a unitary operator, then $\mathcal{F}(o)$ is unitary as well.
\item
If $o$ is an ortogonal projector, then $\mathcal{F}(o)$ is an ortogonal projector too.
\item
$\mathcal{F}(o)$ is non-negative if $o$ is a non-negative operator.
\item
If $o$ is an (essentially) selfadjoint operator, then $\mathcal{F}(o)$ is essentially selfadjoint.
\end{itemize}
The \emph{tangential Fock  functor} $d\Gamma$ assigns to  the operator $o$ an 
operator $\dot{\mathcal{F}}(o)$   defined on $\mathcal{F}(\mathcal{H}')$  by
\begin{equation}
\dot{\mathcal{F}}(o)F_{0}=0,\quad
\dot{\mathcal{F}}(o)[\mathcal{F}^{+}(\phi')^{n}F_{0}]=n\mathcal{F}^{+}(o\phi')\mathcal{F}^{+}(\phi')^{n-1}F_{0}.
\end{equation}
Thus
\begin{itemize}
\item
If  the commutator $[o_{2},o_{1}]$ exists on a dense domain in $\mathcal{H}$,  then
$\dot{\mathcal{F}}([o_{2},o_{1}])=[\dot{\mathcal{F}}(o_{1},
\dot{\mathcal{F}}(o_{2})]$.
\item
If $o\geq 0$, then $\dot{\mathcal{F}}(o)\geq 0$.
\item
If $o$ is an (essentially) self-adjoint  operator,  then $\dot{\mathcal{F}}(o)$  is  
essentially self-adjoint  in $\mathcal{F}$.
\item
If  $o$ generates a strong (semi)group $\exp(-to)$ with real parameter $t$, 
then $\dot{\mathcal{F}}(o)$ generates the strong (semi)group $\mathcal{F}(\exp(-to))$.
\end{itemize}

\subsection{Functional Fock representations}

\subsubsection{Integration on $\Re\mathcal{H}$}

Let $p$ denote an  orthogonal projector in  $\mathcal{H}$
of finite rank $r(p)$.
We assume that  $p$ commute with the conjugation.
Then $p$ is the orthogonal projector of $\Re\mathcal{H}$ onto
$\Re p\mathcal{H}$ as well.

  \smallskip
The \emph{functional integral}  $\int\! d\xi\, F(\xi)$
of a functional $F$ on $\Re \mathcal{H}$ is
   the limit  of the normalized Lebesgue integrals over
the finite dimensional spaces $p\Re \mathcal{H}$
as $p$  converges to the unit operator $\mathbf{1}$, i.e., for
every $\epsilon > 0$ there exists  $p_{\epsilon}$ such that if
$p\Re \mathcal{H}\supset  p_{\epsilon}\Re \mathcal{H}$ then
  the absolute value\begin{equation}
|(2\pi)^{-r(p)/2}\int\! d(p\xi)\, F(p\xi)-
(2\pi)^{-d(p_{\epsilon})/2}\int\! d(p_{\epsilon}\xi)\,
F(p_{\epsilon}\xi)|<\epsilon.
\end{equation}

The finite-dimensional renormalizations are chosen so that the
     Gaussian functional integral
\begin{equation}
\int \! d\xi\,e^{-\|\xi\|^{2}/2}=1.
\end{equation}
A \emph{flag} $(p_{n})=p_{1}<\ldots<p_{n}<\ldots$ is an increasing 
sequence of orthogonal $*$-projectors such that the union 
$\cup (p_{n}\mathcal{H})$ is dense in $\mathcal{H}$.

\begin{proposition}
For any flag $(p_{n})$
\begin{equation}
\lim_{n\rightarrow\infty}(2\pi)^{-d(p_{n})}\int\!d(p_{n}\xi)\, F(p_{n}\xi)
=\int\!d\xi^{*}d\xi\, F(\xi).
\end{equation}
\end{proposition}
\proof
Since $\cup (p_{n}\mathcal{H})$ is dense in $\mathcal{H}$,
for any   positive $\epsilon$ there exists  a projector  $p_{n}$ that has the same rank as $p_{\epsilon}$ and the (constant) Jacobian of the
orthogonal projection of $p_{\epsilon}\mathcal{H}$ onto
$p_{n}\mathcal{H}$ is within $\epsilon$ from $1$. Now
for any $p_{m} > p_{n}$,  the orthogonal projections from
  $(p_{\epsilon} + p_{m})\mathcal{H}$ onto $p_{m}\mathcal{H}$
  have the same Jacobian. 
  
  Thus the integrals in the left hand side of the equation are within $\epsilon$ from  the integral on the right hand side.
     \qed
\begin{proposition}
The functional  integral has the following properties:
\begin{enumerate}
\item  $\int\! d\xi\, F(\xi)$
is a positive  linear functional on the  space of integrable functionals $G$.
\item The integral over a product Hilbert space is equal to the iterated
functional integrals.
\item Integration by parts: Let  $D_{\eta}F$ denote the 
directional derivative
of $F$ in the direction of $\eta\in\Re\mathcal{H}$.  Then
\begin{equation}
\int\! d\xi \,F(\xi)\,D_{\eta}G(\xi)= - \int\!d\xi \,D_{\eta}F(\xi)\,G(\xi)
\end{equation}
   if   $FG\rightarrow 0 $ as the scalar product
     $\xi\rightarrow\infty$ and both integrals exist.
\item The functional integral is invariant under translations
and orthogonal transformations in $\mathcal{H}$.
\end{enumerate}
\end{proposition}
\proof
These  properties follow directly from the corresponding properties of
finite-dimensional Lebesgue integrals. (For the integration by parts
note that for given $\xi$ in  $\mathcal{H}$
  we may choose the projectors  $p'$ such that $p\xi=\xi$.
   \qed

\subsubsection{Gauss Fock representation on $\Re\mathcal{H}$}

In the \emph{Gauss (or real wave) Fock representation on}  $\Re\mathcal{H}$
(compare with \cite{Friedrichs} and \cite{BSZ} )
\begin{itemize} 
\item
  $\mathcal{F}(\mathcal{H})$ is the Gauss Hilbert space
$\mathcal{G}(\mathcal{H})$ the completion  of the space $\mathcal{L}^{2}(\Re\mathcal{H},e^{-\|\phi\|^{2}/2})$
  of functionals $F=F(\xi),\ \xi\in\Re\mathcal{H}$, with $F^{*}(\xi)=\overline{F(\xi)}$ (complex conjugation) and   
  the Hermitian product
\begin{equation}
F^{*}G=\int\! d\xi e^{-\|\xi\|^{2}/2}F^{*}(\xi)G(\xi);
\end{equation}
\item
  the vacuum vector $F_{0}=1$;
\item  
  the annihilators and creators are
\begin{equation}
\mathcal{F}^{-}(\phi^{*})F(\xi)=\partial^{*}F(\xi),
\mathcal{F}^{+}(\phi)F(\phi)=(-\partial F(\xi)+\xi\phi)F(\xi).
\end{equation}
\end{itemize}
\emph{Occasionally we denote $\mathcal{F}^{-}$ and $\mathcal{F}^{+}$ in $\mathcal{G}$ 
as $\mathcal{G}^{+}$ and $\mathcal{G}^{-}$}.

\subsubsection{Bargmann Fock representation on  $\mathcal{H}$}

  Since  $\mathcal{H}$, as a real Hilbert space,  has been identified 
with $\Re(\mathcal{H}\times\mathcal{H}^{*})$,
the \emph{functional integral} $\int\! d\phi^{*}d\phi\, F(\phi,\phi^{*})$
is defined as the limit of Lebesgue
integrals over finite dimensional $*$-subspaces $p\mathcal{H}$.

Now the normalizing constants are $\pi^{-\mbox{\scriptsize dim}(p)}$ so that
\begin{equation}
\int\! d\phi^{*}d\phi\, e^{-\phi^{*}\phi} =1.
\end{equation}

The Hermitian adjoint \emph{Cauchy-Riemann operators} on $\mathcal{H}$ are
\begin{equation}
\partial_{\zeta}=(1/2)(D_{\Re \zeta}-iD_{\Im \zeta}),\
\partial_{\zeta}^{*}=(1/2)(D_{\Re \zeta}+iD_{\Im \zeta}),
\end{equation}
the former being linear and the latter anti-linear in $\phi$.
\smallskip

A \emph{continuous} functional $F$ on $\mathcal{H}^{\infty}$
  is an \emph{entire functional} if
   $\partial_{\zeta}^{*}F(\phi,\phi^{*})=0$  for all  $\phi$ and
   $\zeta$. Notationally  $F=F(\phi)$.

A \emph{continuous} functional $F$ on $\mathcal{H}^{\infty}$
  is a \emph{anti-entire functional} if
   $\partial_{\zeta}F(\phi,\phi^{*})=0$  for all   $\phi$ and
   $\xi$. Notationally  $F=F(\phi)$.

\smallskip
In  the \emph{Bargmann (or complex wave) Fock representation} on  
$\mathcal{H}$ (see \cite{Bargmann})
\begin{itemize}
\item
the  Fock space  $\mathcal{F}(H)$ is the Bargmann space
$\mathcal{B}(\mathcal{H})$,  the (closed)  subspace  of 
anti-entire functionals
$F=F(\phi^{*})$ in $\mathcal{L}^{2}(\mathcal{H}^{*}\times\mathcal{H},
  e^{-\phi^{*}\phi}d\phi^{*} d\phi)$;
 \item The conjugation $F^{*}(\phi^{*})=\overline{F[(\phi^{*})^{*}]}$
\item  The vacuum  functional  $F_{0}=1$;
\item The annihilation and creation operators are
\begin{equation}
\mathcal{F}^{-}(\zeta^{*})F(\phi^{*})=\partial_{\zeta^{*}}F(\phi^{*}),\
\mathcal{F}^{+}(\zeta)F(\phi^{*})= (\phi^{*}\zeta)F(\phi^{*}).
\end{equation}
\end{itemize}
\emph{Occasionally we denote $\mathcal{F}^{-}$ and $\mathcal{F}^{+}$ in $\mathcal{B}$
as $\mathcal{B}^{+}$ and $\mathcal{B}^{-}$}.

\subsection{Bargmann-Segal transform}
The   \emph{coherent functionals}  $F_{\alpha}$ on $\mathcal{H}$ are
\begin{equation}
F_{\alpha} =\sum_{n=1}^{\infty}\frac{1}{n!}\mathcal{F}^{+}(\alpha)^{n}F_{0},\ 
\alpha\in\mathcal{H}.
\end{equation}
By induction, Fock commutation relations imply
\begin{equation}
[\mathcal{F}^{+}(\alpha)^{m}F_{0}]^{*}[\mathcal{F}^{-}(\beta)^{n}F_{0}]
= \delta_{mn}(\alpha^{*}\beta)^{m},
\end{equation}
so that
\begin{equation}
F_{\alpha}^{*}F_{\beta}=F_{\alpha^{*}\beta},
\end{equation}
Then $F_{\alpha}^{*}F_{\alpha} < \infty$  so that
$F_{\alpha}\in\mathcal{F}$ and
  the correspondence between $\alpha$ and $F_{\alpha}$ is
  one to one.
Note that  in Bargmann space $\mathcal{B}$ the coherent functionals
$F_{\alpha}(\psi^{*})=\exp(\psi^{*}\alpha)$.

The  entire functional $F(\alpha)=F^{*}F_{\alpha}$ of the argument $\alpha\in\mathcal{H}$
is the  \emph{ Bargmann-Segal transform} of $F\in\mathcal{F}$. 

The following   proposition is fundamental (see \cite{Berezin71}):
\begin{proposition}
  The coherent functionals $F_{\alpha},\  \alpha\in\mathcal{H}$,
form a continual orthogonal basis of $\mathcal{F}$ as follows:
\begin{enumerate}
\item
Every  $F\in \mathcal{F}$ has  the weak expansion  in $F_{\alpha}$:
   \begin{equation}
  F(\beta^{*})= \int\!d\alpha^{*}d\alpha\,
   e^{-\alpha^{*}\alpha}e^{\beta^{*}\alpha}\,    F(\alpha^{*}).
\end{equation}
\item If $G ,F\in\mathcal{F}$ then
\begin{equation}
  G^{*}F = \int\!d\alpha^{*}d\alpha\, G^{*}(\alpha)F(\alpha^{*}).
\end{equation}
in particular, $\|F \|^{2} = \int\!d\alpha^{*}d\alpha\,
  |F(\alpha^{*})|^{2}$ so that the Bargmann-Segal  transform is one to one.
\end{enumerate}
\end{proposition}
\proof
 The first part follows from the  weak convergence of functional integrals
\begin{equation}
F^{*} \int\!d\alpha^{*}d\alpha\,e^{\beta^{*}\alpha}\, 
F_{\beta^{*}}(\alpha^{*})=
\int\!d\alpha^{*}d\alpha\,
e^{\beta^{*}\alpha}\,( F^{*} F_{\beta^{*}})(\alpha^{*}).
\end{equation}
By the same token the second part follows from the first. In both
cases the commutation with integration is justified by integration over
finite dimensional subspaces with conjugation in $\mathcal{H}$. \qed

\subsection{ Fock Sobolev scales}

Let $o$ be  a real (i.e., commuting with the conjugation)
selfadjoint non-negative  operator in  $\mathcal{H}$ with the
discrete spectrum $\{\lambda_{k}: k=1,2,\ldots\}$. In particular each 
$\lambda_{k}$ has a finite multiplicity $m_{k}$. Assume that the operator $(1+o)^{-p}$ has finite trace for some $p>0$.

\textsc{examples}:  the   harmonic oscillator operator
$-\nabla^{2}+x^{2}$ in $\mathcal{H}=\mathcal{L}^{2}(\mathbb{R}^{n})$,
 positive  globally hypoelliptic operators in
  $\mathcal{L}^{2}(\mathbb{R}^{n})$ (see \cite{Shubin}),
 Beltrami Laplacians, or, more generally,
positive  elliptic operators in $\mathcal{L}^{2}(M)$ on
compact  Riemann manifolds $M$ (see \cite{Shubin}).

\smallskip
For $s\leq 0$,   denote by $\mathcal{H}^{s}$ the \emph{ Hilbert $*$-space}  
of all $\phi\in \mathcal{H}$ with the Hermitian product
$\phi^{*}(1+o)^{s}\psi$. Its antidual $\mathcal{H}^{-s}$  with respect 
to the basic Hertmitean form $\alpha^*\beta$ is the completion of  
$\mathcal{H}$ with respect to the Hermitian product  $\phi^{*}(1+o)^{-s}\psi$. 

If  $s'>s$, then $\mathcal{H}^{s}$ is a dense subspace 
of  $\mathcal{H}^{s'}$, and the iclusions are continuous. Therefore, 
by definition, the family of the Hilbert $*$-spaces 
$\mathcal{H}^{s},\ -\infty<s<\infty$ is a  \emph{ Sobolev scale} generated by $o$. 

The intersection  $\mathcal{H}^{\infty}=\bigcap_{s}\mathcal{H}^{s}$
is the Frechet space with the topology of simultaneous convergence with respect to all   
Hilbert norms. Since $(1+o)^{-p}$ has finite trace for some $p>0$, 
the space $\mathcal{H}^{\infty}$ is nuclear.

Its antidual  with respect to the basic Hertmitean form $\alpha^{*}\beta$ is the strict
inductive limit (see \cite{RS}, Section V.4)				
$\mathcal{H}^{-\infty}=\bigcup_{s}\mathcal{H}^{s}$,
 a nuclear space again. 
 
 Thus we get a Gelfand  triple
\begin{equation} 
\mathcal{H}^{\infty}\subset\mathcal{H}\subset\mathcal{H}^{-\infty}.
\end{equation}

 Similarly, starting with  the Fock quantized $\mathcal{F}$ and  $\mathcal{F}(o)$ 
instead of $\mathcal{H}$ and $o$, we get the
\emph{Fock scale} of Hilbert spaces  $\mathcal{F}^{s}$ and
the triple (see \cite{KMT} and \cite{BSZ}, Section 7.3)
\begin{equation} 
\mathcal{F}^{\infty}\subset\mathcal{F}\subset\mathcal{F}^{-\infty}.
\end{equation} 

Using $\mathcal{F}$ and $\dot{\mathcal{F}}(o)$ instead of $\mathcal{F}(o)$
  we obtain the \emph{tangential Fock scale}  of the Hilbert spaces
 $\dot{\mathcal{F}}^{s}$ and the triple
\begin{equation}
\dot{\mathcal{F}}^{\infty}\subset\mathcal{F} \subset
\dot{\mathcal{F}}^{-\infty}.
\end{equation}
  Note that the  product  states
$[\prod_{j=1}^{n}\mathcal{F}^{+}(\phi_{j})]F_{0}$  belong to
$\mathcal{F}^{\infty}$ if an only if all $\phi_{j}\in
\mathcal{H}^{\infty}$.

\textsc{example} Consider the Fock representation over $\mathcal{H}=\mathbb{C}^{d}$ 
with the standard complex conjugation: 
\begin{eqnarray*}
& &
\mathcal{F}=\mathcal{L}^{2}(\mathbb{R}^{d}),\
F_{0}=(4\pi)^{-1}e^{-u^{2}/4}\\
& &
\mathcal{F}^{-}(u-iv)F(x)=(xu/2 - \partial_{v})F(x),\
\mathcal{F}^{+}(u+iv)F(x)=(xu/2 + \partial_{v})F(x)
  \end{eqnarray*}
  Let $o=1$. 
Then  (see \cite{Obata}, Section 6.2) $\mathcal{F}^{\infty}(\mathbb{C}^{d})$
consists of all  real analytic functions  $F(x)$ such that
for any $\epsilon>0$
\begin{equation}
e^{(1/4 - \epsilon)x^{2}}F\in\mathcal{L}^{1}(\mathbb{R}^{d}),
\end{equation}
and the Fourier transform $G(z)$ of $e^{- x^{2}/4}F$ satisfies
  \begin{equation}
  |G(z)|\prec \exp[(1/2 - \epsilon)z^{2}]
\end{equation}
for all $z\in\mathbb{C}^{d}$.

\smallskip
On the other hand, $\dot{\mathcal{F}}^{\infty}(\mathbb{C}^{d})$ is Schwartz space
$\mathcal{S}(\mathbb{R}^{d})$ of rapidly decreasing infinitely
differentiable functions on $\mathbb{R}^{d}$ (see \cite{KMT}, p.185).

 \smallskip
Note that, if $\phi\in \mathcal{H}^{\infty}$, then  $\mathcal{F}^{\infty}$ 
and $\dot{\mathcal{F}}^{\infty}$ are
invariant   for 
$\mathcal{F}^{+}(\phi)$ and $\mathcal{F}^{-}(\phi^{*})$.

Also, since $\mathcal{H}^{\infty}$ is invariant for pseudodifferential
operators on $X$ (see \cite{Shubin}, Sections 4.3 and 23.2), they are invariant, 
correspondingly,  for quantized and
tangentially quantized pseudodifferential operators.
\begin{remark}
Under the unitary equivalence of Fock representations,
$\mathcal{F}^{\infty}$ and $\mathcal{F}^{-\infty}$ correspond to $(\mathcal{H}^{\infty})$ 
and $(\mathcal{H}^{-\infty})^{*}$ in
 Hida's white noise calculus \mbox{(see \cite{Obata})}.
 
The spaces $ \dot{\mathcal{F}}^{\infty}$ and
$\dot{\mathcal{F}}^{-\infty}$ correspond to the maximal 
 Kristensen-Mejlbo-Poulsen space and their space of temperate distributions  \mbox{(see \cite{KMT})}.
 
Thus their properties are immediately translated into the corresponding properties of $\mathcal{F}^{\infty}$ 
and $\mathcal{F}^{-\infty}$ and  $ \dot{\mathcal{F}}^{\infty}$ and
$\dot{\mathcal{F}}^{-\infty}$.

In particular,  $\mathcal{H}^{\infty}$  and $\mathcal{H}^{-\infty}$ are nuclear spaces,

However,  the spaces $ \dot{\mathcal{F}}^{\infty}$ and
$\dot{\mathcal{F}}^{-\infty}$ are not nuclear. Still they have the Montel
property: their  closed boinded  subsets are compact. In paticular,
these spaces are reflexive.
\end{remark}

\begin{proposition}
The map of $(\phi,F)$ to $\mathcal{F}^{-}(\phi^{*})F$ is continuous 

\mbox{(a)}\ from $\mathcal{H}^{-\infty}\times\mathcal{F}^{\infty}$ to $\mathcal{F}^{\infty})$ 
(and, by duality, from $\mathcal{H}^{-\infty}\times\mathcal{F}^{-\infty}$ to
 $\mathcal{F}^{-\infty}$);

\mbox{(b)}\ from $\mathcal{H}^{-\infty}\times
\dot{\mathcal{F}}^{\infty}$ to $\dot{\mathcal{F}}^{\infty}$ (and, by duality, from 
$\mathcal{H}^{-\infty}\times\dot{\mathcal{F}}^{\infty}$ to
 $\dot{\mathcal{F}}^{\infty}$). 
\end{proposition}
\proof
The first half  of part (a) follows from  Theorem 4.3.9 in \cite{Obata} for  
annihilators  $G(k_{0,1})$;

The first half  of part (b) from the proof of Theorem 4.3.12 in \cite{Obata} 
for  its annihilators $G(k_{1,0})$. \qed

\section{ Cutoff functional derivatives  operators}

\subsection{Functional derivatives operators}
>From now on we assume that $\mathcal{H}=\mathcal{L}^{2}(X)$, where  $X$ is   either a compact Riemann manifold, or the Euclidean space  $\mathbb{R}^{d}$  with the Riemannian measures  $dx$.

The scaling operators $o$ are correspondingly  Beltrami Laplacian, and  harmonic oscillator operators. 

Then $\dot{\mathcal{H}}^{\infty}$ is,
correspondingly, the space $\mathcal{C}^{\infty}(X)$ of infinitely differentiable 
functions on the compact Riemann manifold $X$, and the Schwartz space 
$\mathcal{S}(\mathbb{R}^{d})$ (see \cite{Shubin}, Section 7 and Section 25).

Since   delta-functions $\delta_{x}=\delta_{x}^{*}$
belong to $\mathcal{H}^{-\infty}$, the operators $\mathcal{F}^{-}_{x}=\mathcal{F}^{-}(\delta_{x})$  are well defined, and, by Proposition 1.4, are continuous  
in  $\mathcal{F}^{\infty}$. 

Let
\begin{equation}
\mathcal{F}^{-}_{x_{[n]}}=
\mathcal{F}^{-}_{x_{1}}...\mathcal{F}^{-}_{x_{n}},\ 
(x_{1}\ldots x_{n}\in X^{n}.
\end{equation}
By Proposition 1.4,  for given $G,F\in \dot{\mathcal{F}}^{\infty}$,  the matrix element  
$G^{*}\dot{\mathcal{F}}^{-}_{x_{[n]}}F$  belongs to $\mathcal{H}^{-\infty}$.

 If a \emph{Wick symbol} $W_{k,l}\in(\mathcal{H}^{-\infty})^{k+l}$
 then $W_{k,l}\Big(F^{*}\dot{\mathcal{F}}^{-}_{x_{[k]}}\dot{\mathcal{F}}^{-}_{y_{[l]}}F\Big)$ is a continuous  bilinear  form on $\dot{\mathcal{F}}^{\infty}$. It defines  a continuous \emph{functional derivatives   operator} $\widehat{W}_{k,l}$ from 
$\dot{\mathcal{F}}^{\infty}$ to $\dot{\mathcal{F}}^{-\infty}$ which heuristically is 
\begin{equation}
\widehat{W}_{k,l}=\int\!dx_{[k]}dy_{[l]}\, W_{k,l}(x_{[k]},y_{[l]})
\dot{\mathcal{F}}^{+}_{x_{[k]}}\dot{\mathcal{F}}^{-}_{y_{[l]}}.
\end{equation}

A finite sum  $\widehat{W}=\sum_{k+l\leq m} \widehat{W}_{k,l}$   is a 
\emph{functional derivatives operator of order} $m$ from  $\dot{\mathcal{F}}^{\infty}$ to $\dot{\mathcal{F}}^{-\infty}$.

\smallskip
A  functional derivatives   operator is \emph{local} if  the distributions  $W_{k,l}=W_{k,l}(x)\delta(X)$, 
where $X$ is identified with  the  submanifold  $\{(x,x,...,x)\}\subset X^{k+l}$, and  
$W_{k,l}(x)\in \mathcal{H}^{-\infty}$. Then
\begin{equation}
\widehat{W}=\int\!dx\,\sum_{k\leq m} W_{k,l}(x)(\mathcal{F}^{+}_{x})^{k}(\mathcal{F}^{-}_{x})^{l}.
\end{equation}
Annihilators $\dot{\mathcal{G}}^{-}(\phi^{*})$ in Gauss Fock representation are directional  derivatives $D_{\phi^{*}}$.

Since   delta-functions $\delta_{x}=\delta_{x}^{*}$
belong to $\mathcal{H}^{-\infty}$ it is possible to
consider the  \emph{functional derivative} 
$D_{x}F(\phi)= D_{\delta_{x}}F(\phi)$. Indeed, by Theorem 4.2.4 from \cite{Obata},  
this directional derivative exists for 
$F\in\mathcal{G}^{\infty}$ and coincides with
 $D_{x}=\mathcal{G}^{-}(\delta_{x})$.
 
 On the other hand,   a translation is not a continuous operator  
in $\dot{\mathcal{G}}^{\infty}$ so that $D_{\delta_{x}}$ does not belong  in this space. 
However, by proposition 1.4, it may be continuously extended as  
$\dot{\mathcal{G}}^{-}(\delta_{x})$. It  is the limit of
a family $D_{\eta}$ as $\eta\in\mathcal{H}^{\infty}$ converge
to $\delta_{x}\in\dot{\mathcal{H}}^{\infty})$. By proposition 1.4, this is a  continuous  operator  in  $\dot{\mathcal{G}}^{\infty}$ denoted again 
as $D_{x}$.
  
By Proposition 1.4,  the Hermitian adjoints of the functional derivatives $D_{x}$,
\begin{equation}
D_{x}^{\dagger}F(\phi)=(-D_{\delta_{x}}+\phi(x))F(\phi)
\end{equation}
are continuous operators in $\dot{\mathcal{G}}^{-\infty}$.

Thus the multiplication with  $\delta_{x}$,  which is the operator
$D_{x}+D_{x}^{\dagger}$, is continuous from 
$\dot{\mathcal{G}}^{\infty}$ to $\dot{\mathcal{G}}^{-\infty}$.

The coherent state quadratic form  $F_{\alpha}^{*}\widehat{W}F_{\beta}$ in 
Bargmann space $\mathcal{B}$  is
\begin{eqnarray*}
& &
F_{\alpha}^{*}\Big[\int\!dx_{[k]}dy_{[l]}\,
W_{k,l}(x_{[k]},y_{[l]})
\prod_{i=1}^{l}\mathcal{B}^{+}(\delta_{y_{i}})\prod_{j=1}^{k}
\mathcal{B}^{-}(\delta_{x_{j}})\Big]
F_{\beta}\\
& &
\int\!dx_{[k]}dy_{[l]}\,
W_{k,l}(x_{[k]},y_{[l]})
\prod_{i=1}^{l}(\mathcal{B}^{-}(\delta_{y_{i}})F_{\alpha})^{*}\prod_{j=1}^{k}
\mathcal{B}^{-}(\delta_{x_{j}})
F_{\beta}\\
& &
\int\!dx_{[k]}dy_{[l]}\, W_{k,l}(x_{[k]},y_{[l]})\int\! d\xi^{*}d\xi\:e^{-\xi^{*}\xi}
\prod_{i=1}^{l}\alpha^{*}(y_{i})e^{\alpha^{*}\xi}
\prod_{j=1}^{k}\beta(x_{j})e^{\xi^{*}\beta(x_{j})}\\
& &
=W_{k,l}(\alpha^{*},\beta)e^{\alpha^{*}\beta},
\end{eqnarray*}
where the \emph{Wick symbol} of $\widehat{W}_{k,l}$
\begin{equation}
W_{k,l}(\alpha^{*},\beta)=\int\!dx_{[k]}dy_{[l]}\, W_{k,l}(x_{[k]},y_{[l]})
\prod_{i=1}^{l}\alpha^{*}(y_{i})\prod_{j=1}^{k}\beta(x_{j})
\end{equation}
 is a continuous holomorhic  polynomial of order $(k,l)$ on 
$\mathcal{H}^ {*\infty}\times\mathcal{H}^ {\infty}$.

A \emph{functional derivatives  operator of order} $n$   is a finite sum of   operators
 $\widehat{W}=\sum_{k+l\leq n}\widehat{W}_{k,l}$ with the Wick symbol 
 $W(\alpha^{*},\beta)=\sum_{k\leq m,l\leq n}W(\alpha^{*},\beta)$.

The correspondence between functional derivatives operators and the Wick symbols is one to one.

The continuous complex analytic  polynomial $W(\alpha^{*},\beta)$ is uniquely defined 
by its Taylor coefficients at the origin $(0,0)$.
Therefore, the correspondence between $W(\alpha^{*},\beta)$ and  the   
restricted Wick symbols $W(\alpha^{*},\alpha)$ is one to one.
The  restricted Wick symbols are continuous (real analytic) polynomials on  $\mathcal{H}^{\infty}$.

Real valued  restricted Wick symbols  are \emph{Hamiltonian functionals}, and the
corresponding operators  are  \emph{Hamiltonian operators}.

\subsection{Cutoff  functional derivatives operators}

A  functional derivatives operator $\hat{H}$ is  a \emph{cutoff}  if its Hamiltonian 
functional $W(\alpha^{*},\alpha)$ has  the (unique) continuous extension from 
$\mathcal{H}^{\infty}$ to  $\mathcal{H}^{-\infty}$. This is equivalent to inclusion  
of the terms 
$W_{k,l}(x_{[k]},y_{[l]})\in (\mathcal{H}^{\infty})^{k+l}$
(see \cite{Obata}, the characterization theorem 3.6.2 ); in particular, 
the polynomial $W(\alpha^{*},\alpha)$ belongs to
$\mathcal{G}(\mathcal{H}^{*}\times\mathcal{H})$.

The  Hamiltonian functionals  and their derivatives $D_{\phi}$ in the directions of $\phi\in\mathcal{G}^{\infty}$ are, actually, integrable with respect to the  Gauss measure on  $\mathcal{H}^{-\infty}$. 

 A cutoff operator $\hat{H}$ is a continuous   operator in $\dot{\mathcal{G}}^{\infty}$. 
Thus it has a  dense domain  in $\mathcal{G}$. Its Hermitian adjoint 
$\hat{H}^{\dagger}$ is also cutoff of the same order  with complex conjugate Wick symbol $\bar{H}$. Thus cutoff operators are closable.  

 A cutoff operator $\hat{H}$ is symmetric on $\mathcal{G}^{\infty}$ if and only if its Hamiltonian functional  is  real-valued.

\begin{theorem}
Any  functional derivatives  operator $\hat{H}$ is the strong 
limit of a sequence of cutoff operators $\hat{H}_{n}$.
\end{theorem}
\proof
It suffices to  consider operators  $\hat{H}=\hat{H}_{k,l}$.
Separately for $X=\mathbb{R}^{d}$ and for $X$, a compact Riemann manifold, we construct 
a sequence  of cutoff Wick symbols $W_{n}$ from   
$(\mathcal{H}^{\infty})^{k+l}$ which converges to $C$ in   
$(\mathcal{H}^{-\infty})^{k+l}$ as $n\rightarrow\infty$.

Then the  cutoff operators  $\hat{H}_{n}$ strongly converge to   $\hat{H}$ in the 
topological operator  space $\mathcal{L}(\mathcal{F}^{\infty},\mathcal{F}^{-\infty})$.

\smallskip
\textsc{case of} $X=\mathbb{R}^{d}$.

Let $\chi,\kappa$ be  non-negative infinitely differentiable functions with compact 
support on $\mathbb{R}^{d}$ such that $\chi(0)=1$ and $\int\!dy\; \kappa(y)=1$.
 
 For every $x\in\mathbb{R}^{d}$ the sequence of  $\kappa_{n,x}(y)= n^{d}\kappa(ny-x)$  
from $\mathcal{S}(\mathbb{R}^{d})$ converges to the delta function 
$\delta_{x}$ in $\mathcal{S}'(\mathbb{R}^{d})$ as $n\rightarrow\infty$.
At the same time the sequence of $\chi_{n}(x)=\chi(x/n)$ converges to $1$ in  $\mathcal{S}'(\mathbb{R}^{d})$ as $n\rightarrow\infty$.

Now  the sequence of the cutoff Wick symbols from $\mathcal{S}(\mathbb{R}^{d})^{k+l}$
\begin{equation}
W_{n}(x_{[k+l]})=\prod_{1}^{ k+l}\chi (x_{i}/n)\int\!\prod_{1}^{ k+l}dy_{i}\; \kappa_{n,x_{i}}(y_{i})c(y_{[k+l]})
\end{equation}
converges to $C(x_{[k+l]})$ in $\mathcal{S}'(\mathbb{R}^{d})^{k+l}$ as $n\rightarrow\infty$.

\smallskip
  \textsc{Case of a compact Riemann manifold} $X$.

In this case $\chi_{n}(x)=1$ for all $x$.

Since the geodesic  exponential mapping is one to one on an open neighborhood $W$ 
of the diagonal in $X\times X$ ,  for every pair $(x,y)\in W$ there is a 
unique geodesic curve  from $x$ to $y$ in $X$. Let $sy$  denote the  point   
at the geodesic distance  $s$ from $x$.

  Choose a non-negative infinitely differentiable  function $\kappa(x,y)$ 
on $X\times X$ with  support in $W$ such that  
  $\int\!dy\; \kappa(x,y)=1$ for all $x$. Let $\kappa_{x}(y)= \kappa(x,y)$.

Then  the sequence of the cutoff Wick symbols
\begin{equation}
W_{n}(x_{[k+l]})=\int\!\prod_{1}^{ k+l}dy_{i}\; \kappa_{n,x_{i}}(y_{i})c(y_{[k+l]})
\end{equation}
belong to  $(\mathcal{H}^{\infty})^{k+l}$ and converges to the  Wick symbol  $W(x_{[k+l]})$ in the topology of $(\mathcal{H}^{-\infty})^{k+l}$ 
as $n\rightarrow\infty$.
\qed

A continuous polynomial 
$A(\phi^{*},\phi)\in
\mathcal{G}(\mathcal{H}^{*}\times\mathcal{H})$ is the \emph{antinormal symbol} 
of $\widehat{W}$ if the coherent state matrix  of $\widehat{W}$ in the 
Bargmann Fock space $\mathcal{B}$
\begin{equation}
F_{\alpha}^{*}\widehat{W}F_{\beta}=\int\! d\phi^{*}d\phi\; e^{-\phi^{*}\phi}e^{\alpha^{*}\phi}
A(\phi^{*},\phi)e^{\phi^{*}\beta}.
\end{equation}
The functional $e^{\alpha^{*}\phi}$ of $(\alpha^{*},\phi)$ is the integral kernel of  
the identity operator on the closed Bargmann subspace  $\mathcal{B}$  of anti-entire  
functionals $A(\phi^{*})$  in the Gauss Hilbert space $\mathcal{G}$, and is orthogonal 
to all entire functionals $E(\phi)$. Therefore  
$e^{\alpha^{*}\phi}$ is the integral kernel of  the orthogonal projector $\mathbf{P}$ 
of  $\mathcal{G}$ onto $\mathcal{B}$.

\begin{theorem}
Let $\widehat{W}$ be a local cutoff   functional derivative operator.

 If the  Hamiltonian functional  $W(\alpha^{*},\alpha)$ is  bounded from below on 
$\mathcal{H}^{\infty}$ then the  Hamiltonian operator $\widehat{W}$ is  
lower bounded on $\mathcal{B}^{\infty}$. 
\end{theorem}
\proof
The Hamiltonian functional
\begin{eqnarray*}
& &
W(\alpha^{*},\alpha)=e^{-\alpha^{*}\alpha}\int\!d\phi^{*}d\phi\;
e^{-\phi^{*}\phi + \alpha^{*}\phi+\phi^{*}\beta}A(\phi^{*},\phi)\\
& &
=\int\!d\phi^{*}d\phi\;e^{-(\alpha^{*}-\phi^{*})(\alpha-\phi)}
A(\phi^{*},\phi)
\end{eqnarray*}
The Poisson transformation semigroup
\begin{eqnarray*}
& &
W(\alpha^{*},\alpha;t)=\int\!d\phi^{*}d\phi\,e^{-(\alpha^{*}-\phi^{*})(\alpha-\phi)/t}A(\phi^{*},\phi)\\
& &
\int\!d\phi^{*}d\phi\,e^{-\phi^{*}\phi/\sqrt{t}}
A(\phi^{*}+\sqrt{t}\alpha^{*},\phi+\sqrt{t}\alpha)
\end{eqnarray*}
is the fundamental solution for the diffusion equation
\begin{equation}
(\partial_{t}-\Delta_{G})W(\alpha^{*},\alpha;t)=0,\ t > 0,\ 
W(\alpha^{*},\alpha;0+)=A(\alpha^{*},\alpha),
\end{equation}
where $\Delta_{G}=\int\!dx\,D_{x}^{2}$
is the  \emph{Gross Laplacian} (see \cite{Obata}, Section 5.3).

By theorem 5.2.5 from \cite{Obata}, the Poisson group is a strongly continuous 
operator semigroup in $\mathcal{G}^{\infty}$
generated by $\Delta_{G}$. Note that  antinormal symbols of all cutoff Hamiltonian  operators belong to  $\mathcal{G}^{\infty}$.

The Gross Laplacian   maps continuously $\mathcal{G}^{\infty}$  into $\mathcal{G}^{\infty}$   
(see \cite{Obata}, Proposion 5.3.2) and, therefore, has a dense domain in $\mathcal{G}$ 
which includes  antinormal symbols of all cutoff Hamiltonian operators.

All above shows that the Hamiltonian functional
\begin{equation}
W(\alpha^{*},\alpha)=e^{\Delta_{G}}A(\alpha^{*},\alpha)\sum_{n\geq 0}[(-1)^{n}/n!]\Delta_{G}^{n}A(\alpha^{*},\alpha),
\end{equation}
the latter series being  just a finite sum, justifying the heuristic
expression for $e^{\Delta_{G}}$. Now the formal inversion makes sense
\begin{equation}
A(\alpha^{*},\alpha)=e^{-\Delta_{G}}A(\alpha^{*},\alpha)=\sum_{n\geq 0}[(-1)^{n}/n!]\Delta_{G}^{n}W(\alpha^{*},\alpha)
\end{equation}
In particular, any cutoff operator $\widehat{W}$
has a unique antinormal symbol $A$;   that the polynomials 
$W(\alpha^{*},\alpha)$ and $A(\alpha^{*},\alpha)$ have the same  order; 
and that  the order of the polynomial  $W(\alpha^{*},\alpha)-A(\alpha^{*},\alpha)$ 
is strictly less than the order  of the polynomial 
$A(\alpha^{*},\alpha)$.
Since the lower bound of  the operator  $\widehat{W}$ is never less than the lower bound 
of its antinormal symbol  $A$, this completes the proof. \qed

\subsection{Antinormal Feynman integral}
By Theorem 2.2,  a cutoff operator $\hat{H}$ with the lower bounded Hamiltonian  functional $H$   has the Friedrichs extension from $\mathcal{H}^{\infty}$. Let us  preserve the notation $\hat{H}$  for this extension.
\begin{theorem}
Let $A(\phi^{*},\phi)$  be   the antinormal symbol of a cutoff operator $\hat{H}$.

Then the coherent state matrix
 $F_{\alpha}^{*}e^{-i\hat{H}}F_{\beta}$ is equal to
\begin{equation}
\lim_{N\rightarrow \infty}
\int\prod_{j=1}^{N}\!d\phi_{j}^{*}d\phi_{j}\,
\exp\sum_{j=0}^{N}\Big[(\phi_{j+1}-\phi_{j})^{*}\phi_{j} -
iA(\phi_{j}^{*},\phi_{j})/N\Big]
\end{equation}
with $\phi_{N+1}= \alpha,\ \phi_{0} = \beta$.
\end{theorem}
\proof
As in \cite{KS}, pp. 69-70,  consider  the strongly differentiable  operator family in $\mathcal{B}$
\begin{equation}
[O(t)F](\alpha^{*})= \int\!d\phi^{*}d\phi\, e^{-\phi^{*}\phi}
e^{\alpha^{*}\phi}e^{-iA(\phi^{*},\phi)t}F(\phi^{*})
\end{equation}
We have  $\|O(t)\|\leq 1$ (since $|e^{-iA(\phi^{*},\phi)t}|=1$), and
the strong $t$-derivative $A'(0)=\hat{H}$.  Then, by the Chernoff's product theprem \cite{Chernoff}, the evolution operator 
\begin{equation}
e^{-i\hat{H}}F=\lim_{N\rightarrow \infty}[H(1/N)]^{N}F.
\end{equation}

\smallskip
The coherent state  matrix   $F_{\alpha}^{*}[A(t/N)]^{N}F_{\beta}$ is 
the $N$-iterated Gaussian integral over $\mathcal{H}$ which, by the Fubini's theorem, 
is equal to  the $N$-multiple  Gaussian integral over $\mathcal{H}^{N}$. \qed

\begin{remark}
In the notation $\tau_{j}=jt/N,\ \phi_{\tau_{j}} = \phi_{j},\ 
j=0,1,2,\ldots,N$,
and  $\Delta \tau_{j}= \tau_{j+1}-\tau_{j}$, the  multiple integral (2) is
\begin{equation}
\int\prod_{j=1}^{N}\!d\phi_{\tau_{j}}^{*}d\phi_{\tau_{j}}\:
\exp i\sum_{j=0}^{N}\Delta t_{j}\left[-i(\Delta\phi_{\tau_{j}}/
\Delta \tau_{j})^{*} \phi_{\tau_{j}}\rangle - 
A(\phi_{\tau_{j}}^{*},\phi_{\tau_{j}})\right].
\end{equation}
 Its limit at $N=\infty$  is a rigorous mathematical  definition 
 of the heuristic Hamiltonian  Feynman type integral over histories, with the  
higher derivatives renormalization  $A$ of the Hamiltonian functional $H$,  
for the coherent state matrix
\begin{equation}
\int_{\alpha}^{\beta}\prod_{0< \tau < t}\!d\phi_{\tau}^{*}d\phi_{\tau}\:
 \exp i\int_{0}^{t}d\tau
\left[-i(\partial_{\tau}\phi_{\tau})^{*} \phi_{\tau} - 
A(\phi_{\tau}^{*},\phi_{\tau})\right].
\end{equation}
\end{remark}

\section{Quantized Galerkin approximations}

Let $\{p_{n}\}$ be a flag of \emph{finite dimensional} orthogonal projectors 
in $\mathcal{H}^{\infty}$ (so that $ p_{n}$ is an 
increasing sequence of  orthogonal projectors strongly converging to the 
unit operator on a dense subspace in $\mathcal{H}$).

 Then $\{P_{n}=\mathcal{G}(p_{n})\}$ is the corresponding flag of 
infinite dimensional  \emph{quantized } orthogonal projectors in $\mathcal{G}$. 

\smallskip
For  a cutoff Hamiltonian operator  $\hat{H}$ in the Gauss space $\mathcal{G}$,  the
\emph{reduced Hamiltonian operators } $\hat{H}_n$   are 
Friedrichs extensions of $P_{n}\hat{H}P_{n}$   in  $\mathcal{G}$. They are uniformly bounded  from below.

\begin{proposition}
Reduced Hamiltonian operators  $\hat{H}_n$ are
polynomial partial differential operators in $P_{n}\mathcal{G}$ with the 
normal symbol $H(p_{n}\alpha^{*},p_{n}\alpha)$.
\end{proposition}
\proof
Coherent state matrix  elements of $\hat{H}_n$ are
$F^{*}_{\alpha}P_{n}\hat{H}P_{n}F_{\beta}=F^{*}_{p_{n}\alpha}\hat{H}F_{p_{n}\beta}$. \qed

\medskip

Let $f$ be a  complex bounded continuous function  on the  real axis 
$\mathbb{R}^{+}$. 
Then, by the spectral theorem, for any selfadjoint non-negative operator $T$ in  
$\mathcal{G}$ the operator $f(A)$ is bounded with the operator norm $\leq \sup |f|$.  
If a family of such functions $f_{t}$ depends continuously on a parameter $t$ 
in a compact $K\subset\mathbb{R}$ then the operator family $f_{t}(A)$ is 
uniformly strongly continuous on $K$ with respect to $t$.

\begin{theorem}
The operators $f_{t}(\hat{H})$ are strong operator limits of the 
operators $f_{t}(\hat{H}_n)$ as $n\rightarrow\infty$, uniformly on compact  
$t\geq 0$-intervals.
\end{theorem}
\proof
\textsc{part i}\
The sequence   $\hat{H}_nF$ converges strongly to $\hat{H}F$ in   $\mathcal{G}$.

\proof  
Since the cutoff operator $\hat{H}$  is continuous
in $\mathcal{G}^{\infty}$, the bilinear form $G^{*}\hat{H}F$ is 
separately continuous on that Frechet space. By a Banach theorem
(see \cite{Rudin}, Theorem 2.17), the bilinear form is, actually continuous 
on $\mathcal{G}^{\infty}$. Along with the equality
\begin{equation}
P_{n}F(\phi^{*},\phi)=F(p_{n}\phi^{*},p_{n}\phi),
\end{equation}
this implies that the  operator $\hat{H}$ is the weak limit of
$\hat{H}_n=P_{n}\hat{H}P_{n}$ in $\mathcal{G}^{\infty}$.
Since   $\mathcal{G}^{\infty}$ is a  Montel space,  a weakly covergent sequence 
$P_{n}\hat{H}P_{n}F$ converges in  the topology of $\mathcal{G}^{\infty}$, and, therefore, of $\mathcal{G}$. \qed

\smallskip
\textsc{part ii}\
If  $\lambda$ is a given  complex number with non-zero imaginary part,
 then, for any $G\in\mathcal{G}$, the sequence of the resolvents
$(\lambda-\hat{H}_n)^{-1}G$ converges strongly to 
$(\lambda-\hat{H})^{-1}G$. 
\proof
Since the operator norms $\|(\lambda+\hat{H}_n)^{-1}\|$ are uniformly bounded, 
it suffices to consider the dense set  of
$G=(\lambda+i\hat{H})^{-1}F$ with $F\in\mathcal{G}^{\infty}$. In such a case
\begin{eqnarray*}
& &
\|(\lambda+\hat{H}_n)^{-1}G-(\lambda+\hat{H})^{-1}G\|\\
& &
= \|(\lambda+\hat{H}_n)^{-1})(\hat{H}_n-\hat{H})
(\lambda+\hat{H})^{-1}F\|\\
& &
\prec |\Im\lambda|^{-1}\|(\hat{H}_n-\hat{H})
(\lambda+\hat{H})^{-1}F\|,
\end{eqnarray*}
which converges to zero, by the \textsc{part i}.

\smallskip
\textsc{part iii}\   As in the proof of theorem VIII.20 in \cite{RS}, the 
\textsc{part ii} implies the strong convergence  of  $f_{t}(\hat{H}_n)$ to
$f_{t}(\hat{H})$ (uniformly on compact  $t$-intervals).
\qed

\begin{corollary}
The sequence $e^{-i\hat{H}_nt}$ converges strongly  to
$e^{-i\hat{H}t}$ as $N\rightarrow\infty$, uniformly on compact $t$-intervals.

 In particular, any solution $F(\phi^{*},\phi;t)$ of the corresponding functional 
derivatives Schr\"{o}dinger equation
\begin{equation}
\partial_{t}F+i\hat{H}F=0,\  F(\phi^{*},\phi;0) \in\mathcal{D}(\hat{H})
\end{equation}
 is the limit of the solutions $F_{n}\in\mathcal{D}(\hat{H}_n)$ as $n\rightarrow\infty$ 
of the partial differential  Schr\"{o}dinger equations
\begin{equation}
\partial_{t}F_{n}+i\hat{H}_nF_{n}=0,\ F_{n}(\phi^{*},\phi;0)
=P_{n}F(\phi^{*},\phi; 0)\in\mathcal{D}(\hat{H}_n)
\end{equation}
uniformly on compact $t$-intervals.
\end{corollary}
\begin{remark}
 Theorem 2.3 (applied  to $\mathcal{H}_{n}=p_{n}\mathcal{H}$)  shows that  
the antinormal   Feynman integral for the reduced Schr\"{o}dinger equation is 
the limit of multiple finite dimensional integrals with respect to Gaussian measures. 

This is a (convergent!)   alternative for standard space-time lattice 
approximations in quantum field theory.
\end{remark}
 In a forthcoming  paper  we show that  the rate of convergence is $\prec t^{2}/n$ 
so that the limit exists  if $t\rightarrow \infty$  with the rate  
$n^{(1+\epsilon)/2},\epsilon >0$. Therefore, the remark is applicable to scattering matrices.

\bibliographystyle{amsalpha}

\begin{thebibliography}{A}
\bibitem [1]{BSZ}
Baez, J.C., Segal I.E., Zhou Z.: \textit{Introduction to Algebraic and Constructive 
Quantum Field Theory}, Princeton University Press, 1992.
 
 \bibitem [2]{Bargmann}
 Bargmann, V.: \textit{Remarks on a Hilbert space of analytic functions},
Proc. National Acad. Sci. USA, \textbf{48} (1962), 199-204.

 \bibitem [3]{Berezin}
 Berezin, F.A.: \textit{The Method of Second Quantization},
 Academic Press, 1966.

 \bibitem [4]{Berezin71}
Berezin, F.A.: \textit{Wick and anti-Wick symbols of operators,
Math. USSR Sbornik} \textbf{15} (1971), 577-606.

\bibitem [5]{BS}
Bogoliubov, N .N.,   Shirkov, D. V.:
\textit{Introduction to the Theory of Quantized Fields}, Wiley,  1980.

\bibitem [6]{Chernoff}
Chernoff, P.: Note on product formulas for operator semigroups,
\textit{J. Func. Analysis} \textbf{2} (1968), 238-242.

\bibitem [7]{D}
Dynin, A., \textit{Feynman integral for functional Schr\"{o}dinger equations},  Amer. Math. Soc. Transl. Ser. 2, \textbf{206} (2002), 65-80.

 \bibitem[8]{FS}
 Faddeev, L.D., Slavnov A.A.:  \textit{ Gauge Fields: An Introduction to 
 Quantum Theory}, Addison-Wesley (Frontiers in Physics, vol. 83), 1991.
 
 \bibitem [9]{Friedrichs}
 Friedrichs, K.O.: \textit{Mathematical Aspects of the Quantum Theory of Fields},  
Interscience, 1953.

\bibitem [10]{GJ}
 Glimm J., Jaffe, A.: \textit{Quantum Physics:  a Functional Integral Point of View}, 
Springer-Verlag, 1987.

\bibitem [11]{KS}
Klauder, J.,R., Skagerstam, B.: \textit{Coherent States},
World Scientific,  1985.

\bibitem [12]{KMT}
Kristensen, P., Melbo, l.,  Thue Poulsen, E.:
Tempered distributions in infinitely many dimensions,
\textit{Commun. Math. Phys. }(1965),\textbf{1}, 175-214.
 

\bibitem [13]{Obata}
Obata, N.: \textit{White Noise Calculus and Fock Space},
Lecture Notes in Mathematics, no.1557, Springer-Verlag, 1994.

\bibitem [14]{Shubin}
M. A. Shubin, \textit{Pseudodifferential Operators and Spectral
Theory}, 2nd edition, Springer, 2001.


\bibitem [15]{RS}
Reed, M., Simon B.: \textit{Methods of Modern Mathematical Physics}, 
Academic Press,, vol.I, 1972;  vol. II, 1975.

\bibitem [16]{Rudin}
 Rudin, W.: \textit{Functional Analysis}, McGGraw-Hill, , 1973.
\end{thebibliography}

\end{document}